\documentclass[twocolumn,showpacs,preprintnumbers,amsmath,amssymb,epsf,superscriptaddress]{revtex4-1}

\usepackage{graphicx,epsfig}
\usepackage{dcolumn}
\usepackage{bm}
\usepackage{color}
\usepackage{comment}

\begin{document}


\title{Characteristic Sensitivity of Turbulent Flow within a Porous Medium under Initial Conditions}
\author{Takehito Suzuki}
\email{t-suzuki@phys.aoyama.ac.jp}
\affiliation{Department of Physics and Mathematics, Aoyama Gakuin University, 5-10-1 Fuchinobe, Chuo-ku, Sagamihara, Kanagawa 252-5258, Japan}


\begin{abstract}
Flows within porous media play important roles in many scientific and industrial systems. However, the case wherein such flows become turbulent has not been completely understood, particularly, from a mathematical viewpoint. In this study, the $k-\varepsilon$ model (the variable $k$ denotes the turbulent kinetic energy per unit mass and $\varepsilon$ the dissipation rate for the turbulent kinetic energy), which has been widely used for the usual turbulent flow in a clear fluid, was applied to a turbulent flow through porous media.  If a homogeneous, averaged flow and homogeneous isotropic turbulence are assumed, the governing equations for the variables $k$ and $\varepsilon$ describe a straight line as a nullcline common to both the variables. The nullcline was shown to be a line attractor. A temporal evolution of the eddy viscosity was also obtained, and the initial and final values of the eddy viscosity were observed to be related to a power law, indicating universal sensitivity. This sensitivity originates from the common nullcline and is not observed for the usual turbulent flow. Finally, nonlinear mathematical and seismological implications were provided using based on the results obtained.
\end{abstract}

\maketitle

\section{Introduction} \label{secIn}

Porous media, or those media with a solid skeleton and many pores inside, are substantially studied in several scientific and industrial fields, including oil reservoirs, rockfill dams, fluidized bed combustion, biomaterials, filtration membranes, and chemical wastes \cite{Zhang, Kho, Gar, Kar, Kle, LiJ}. For these aforementioned cases, the pores are considered to be filled with fluid, i.e., water and/or oil. A particular topic of research interest in these fields seems to be directed at the fluid flow and transport phenomena \cite{Ami, Tah, Men, Sha}. The most commonly adopted principle for many studies on these topics is the Darcy law, which states that flow rate through a porous medium is proportional to the fluid pressure gradient. This law assumes a laminar flow.

Porous media have also been of interest in seismological studies \cite{Biz06a, Biz06b, Ric06, Suz06, Suz09, Sega, Suz14, Suz17} mainly because fault rocks characteristically contain pores, and fluid (water) is considered to fill these pores. Particularly, an increase (or decrease) in the fluid pressure induces decrease (or increase) in the effective normal stress that acts on the fault zone, thereby decreasing (or increasing) friction stress and increasing (or decreasing) slip velocity. Many processes, such as thermal pressurization and dilatancy, are the presumed mechanisms causing fluid pressure variation \cite{Biz06a, Biz06b, Ric06, Suz06, Suz09, Sega, Suz14, Suz17}. Additionally, the increase (decrease) in the fluid pressure in the fault zone induces the fluid outflow to (inflow from) the outside of the fault zone because of the emergence of the gradient of the fluid pressure profile along the direction normal to the fault zone. This outflow (inflow) also reduces (raises) the fluid pressure in the fault zone, thereby affecting the slip behavior \cite{Suz06, Suz09, Sega, Suz14}.

In fluid mechanics, turbulent flow in a clear fluid (the fluid in the system consisting of only the fluid phase, not the fluid in porous media) has attracted significant research interest. In actuality, two main problems exist for the treatment of the turbulent flow. The first is the complexity involved in exactly simulating a real flow field because the degree of freedom for the flow is significantly high. The second is the sensitive dependence of the system behavior on the initial conditions, as widely known for turbulent flows; i.e., the Navier$-$Stokes equation has sensitive dependence on the initial conditions \cite{Tak}. The viewpoint of  time average or ensemble average has been introduced to circumvent the aforementioned complexity (i.e., the first problem) and sensitivity (i.e., the second problem). Many frameworks using such averages have been constructed; these frameworks include the mixing length model \cite{Sch, Oer}, $k-\varepsilon$ model (where $k$ denotes the turbulent kinetic energy per unit mass and $\varepsilon$ the dissipation rate of $k$ into a larger wave-number space) \cite{Lau74, Yak, Bes, Shi, Zhao, Lin, Fen}, and the Reynolds stress transport model \cite{Lau75, Gib, Jac}. Among these models, the $k-\varepsilon$ model has been widely used. This model requires the decomposition of the fluid velocity field into the time-averaged velocity and deviation from it. The variables $k$ and $\varepsilon$ are expressed in terms of the deviation and used to calculate the eddy viscosity $\nu_T$, which is a quantitative measure of how momentum is transported from the time-averaged velocity field to the deviation thereof. 

Turbulent flow in porous media has also been widely studied from theoretical \cite{Hub, Mas, Ant, Ven, Get, deL01, Ped01, Hsu, Brag, Chand06, ChanH07, deL03, Ped03, Lie, Sai, Nak08, Ped08, Kum, Ter, Nim, Jou15, Jou17, HeX} and experimental \cite{Segu, Nak07} viewpoints. Of the models that treat turbulent flow, the $k-\varepsilon$ model has been widely employed \cite{Mas, Ant, Get, deL01, Ped01, Hsu, Brag, Chand06, ChanH07, Ped03, Lie, Sai, Nak08, Ped08, Kum, Ter, Nim, Jou15, Jou17}. To consider the temporal evolutions of $k$ and $\varepsilon$ in porous media, it should be noted that energy is supplied from the time-averaged velocity to its deviation via pore walls because the averaged flow creates disturbance upon hitting the pore walls. Therefore, finite $k$ and $\varepsilon$ values in the steady state with constant time-averaged flow can exist in porous media \cite{Ped01}, which is not the case for clear fluids. Such a statement implies that there can exist infinitely many steady states; i.e., attractors are continuously distributed on the $k-\varepsilon$ phase space. Obtaining solution orbits connecting the initial states and attractors is a useful step to understanding the temporal evolution of the turbulent flow in porous media. Particularly, the sensitive dependence of the solution orbits on the initial conditions is expected by analogy with a previous nonlinear mathematical treatment \cite{Suz17}, which revealed that a continuous attractor can produce sensitivity to the initial conditions. 

Previous studies assumed variations in the initial conditions for $k$ and $\varepsilon$ in porous media. For example, several researchers assumed the value $k_0/\varepsilon_0$ (where $k_0$ and $\varepsilon_0$ denote the initial values of $k$ and $\varepsilon$, respectively) to be constant \cite{Yak, Shi}. Another study \cite{Hsu} employed the ``constant eddy viscosity'' condition that was initially introduced for clear fluids \cite{Lin}. Lin and Liu \cite{Lin} introduced a small amount of $k_0$ in the initial condition as a seed for generating turbulence in clear fluids, whereas $\varepsilon_0$ was determined by the condition that the eddy viscosity was constant. We should systematically treat these conditions.

Continuous attractors have drawn the interest of many researchers, especially those from the field of neural networks \cite{Sam, Seu, YuJ, Mil, Zhan}. For example, line attractors have been explored for studying oculomotor control. However, the mathematical understanding of such systems, particularly concerning the effect of the disturbance in the initial conditions on the steady state, has not been performed thus far. To understand this effect, the analogy of a turbulent flow in porous media works well.

This paper is organized as follows. Our model and its governing equation system are presented in Sect. \ref{secMS}. A homogeneous, time-averaged flow and a homogeneous, isotropic disturbance are assumed, and the analytical treatment for the state is discussed in Sect. \ref{secHSAT}. In this section, the universal sensitivity of the final state to the initial state is emphasized for the turbulent flow in porous media. Moreover, the sensitivity of the final eddy viscosity on the initial eddy viscosity, and that of the difference in the final eddy viscosity on the difference in the initial eddy viscosity are clarified. The findings in this study, as well as some nonlinear mathematical and seismological implications, are summarized in Sect. \ref{secDisCon}.

\section{Mathematical Formulation} \label{secMS}

We now present a mathematical framework that governs the behavior of turbulent fluid flow in a porous medium. Let us consider an infinite porous medium, whose pores are filled with fluid, i.e., water. The solid phase of the medium is assumed to be rigid such that the porosity $\phi$ remains constant. The fluid phase is assumed to be single-phase Newtonian fluid and incompressible such that the density of the fluid phase $\rho$ remains constant. Moreover, consider the pore Reynolds number $\mathrm{Re}_p$, expressed as $\mathrm{Re}_p \equiv d v_c/\nu$, where $d$, $v_c$, and $\nu$ denote the characteristic pore scale, characteristic flow velocity scale, and viscosity of the fluid phase, respectively. Reportedly \cite{Ped01, Dyb}, a flow can be considered turbulent when 
\begin{equation}
\mathrm{Re}_p > \mathrm{Re}_c,  \label{eqRep}
\end{equation}
where $\mathrm{Re}_c$ denotes the critical pore Reynolds number, above which the flow becomes turbulent. Generally, $\mathrm{Re}_c$ is considered to be in the range of $300-1000$ \cite{Ant, Nak08, Dyb}.

To deal with such turbulent flow within porous media, we utilize the $k-\varepsilon$ model constructed by \cite{Mas, Ped01}. First, let us suppose that the length scale is smaller than the characteristic pore scale, i.e., from a microscopic viewpoint. The framework treating the ``usual turbulent flow,'' or the ``turbulent flow in a system consisting of only the fluid phase (clear fluid)'' can be used for this length scale. We write $k$ (turbulent kinetic energy) and $\varepsilon$ (dissipation rate of $k$ into a large wave-number space) for the usual turbulent flow as $k^{\mathrm{usual}}$ and $\varepsilon^{\mathrm{usual}}$, respectively, and the mathematical framework of the $k-\varepsilon$ model in porous media is constructed from $k^{\mathrm{usual}}$ and $\varepsilon^{\mathrm{usual}}$ in the following discussion.

We then decompose the flow velocity profile into time-averaged velocity and deviations from it, as mentioned in Sect. \ref{secIn}. The time average of variable $A$, $\overline{A}$, is defined as
\begin{equation}
\overline{A} \equiv \frac{1}{\Delta t} \int_t^{t+\Delta t} A dt, \label{eqTAv}
\end{equation}
where $\Delta t$ denotes the integration time interval. Performing this integration eliminates the high-frequency components of the variable. Using Eq. (\ref{eqTAv}), the velocity field $\bm{u}$ is expressed as
\begin{equation}
\bm{u}=\overline{\bm{u}}+\bm{u}', \label{eqAD}
\end{equation}
where $\bm{u}'$ is the deviation of $\bm{u}$ from $\overline{\bm{u}}$. Notably, $\overline{\bm{u}'}=\bm{0}$ by definition.

For the usual turbulent flow, $k^{\mathrm{usual}}$ and $\varepsilon^{\mathrm{usual}}$ can be written in terms of $\bm{u}'$ as \cite{Lau75, Yak}
\begin{equation}
k^{\mathrm{usual}}=\frac{ \overline{ \bm{u}'^2 } }{2}, \label{eqdef-k}
\end{equation}
and
\begin{equation}
\varepsilon^{\mathrm{usual}}=\nu \overline{ \frac{\partial u'_i}{\partial x_j} \frac{\partial u'_j}{\partial x_i}  }, \label{eqdef-e}
\end{equation}
where $i,j=1,2,3$ and the repeated indices are summed. Equation (\ref{eqdef-k}) is based on the following assumption: 
\begin{equation}
(\overline{ \bm{u}' \bm{u}' })_{ij} = \overline{ u'_i u'_j } =-\nu_T \left( \frac{\partial \overline{u}_i}{\partial x_j} + \frac{\partial \overline{u}_j}{\partial x_i} \right) +\frac{2}{3} k \delta_{ij}, \label{eqdef-Rs}
\end{equation}
where $\nu_T$ denotes the eddy viscosity and $\delta_{ij}$ the Kronecker delta. The term $ \overline{ \bm{u}' \bm{u}' } $ is known as the Reynolds stress, and Eq. (\ref{eqdef-Rs}) shows that the momentum is transported from the time-averaged velocity $\overline{\bm{u}}$ to its deviation $\bm{u}'$. Equation (\ref{eqdef-Rs}) also states that the Reynolds stress is directly proportional to the local time-averaged velocity gradients and that the proportionality factor, i.e., the eddy viscosity $\nu_T$, is a scalar quantity. Notably, $\nu_T$ quantitatively characterizes how the flow is disturbed and is regarded herein as the intensity of disturbance. 

\begin{figure}[tbp]
\centering
\includegraphics[width=8.5cm]{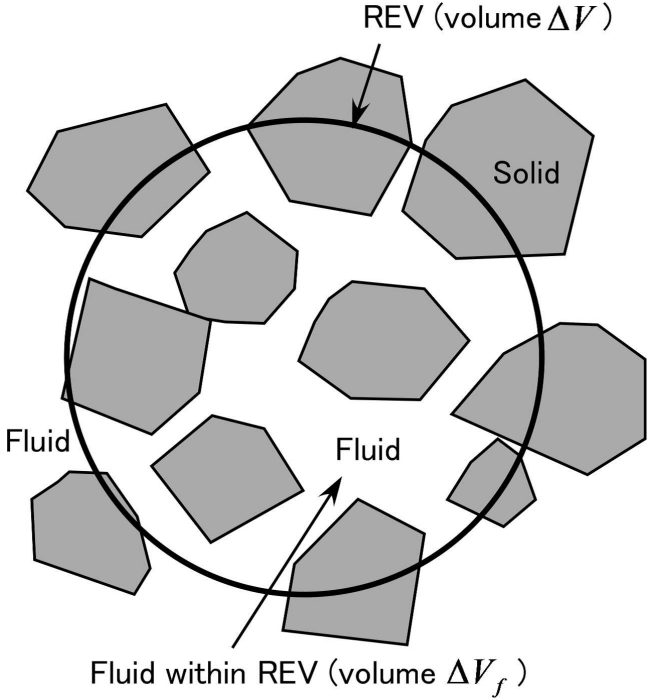}
\caption{Integration range for the spatial average. The integration is performed within $\Delta V_f$.}
\label{FigREV}
\end{figure}

Nonetheless, we must consider a solid skeleton while discussing $k$ and $\varepsilon$ in porous media. Thus, we must consider a coarse graining viewpoint associated with the representative elementary volume (REV), as shown in Fig. \ref{FigREV}. The volume of REV, $\Delta V$, is several times larger than the characteristic pore scale and significantly smaller than the system size; thus, it becomes necessary to take the spatial averages of the variables within the REV. Accordingly, the spatial average of variable $A$, $\langle A \rangle$, is expressed as
\begin{equation}
\langle A \rangle \equiv \frac{1}{\Delta V_f} \int_{\Delta V_f} A dV, \label{eqOp1}
\end{equation}
where $\Delta V_f$ denotes the volume of the fluid phase within $\Delta V$. We, therefore, conclude that the temporal evolutions of $\langle k^{\mathrm{usual}} \rangle$ and $\langle \varepsilon^{\mathrm{usual}} \rangle$ must be obtained to understand the behavior of the turbulent flow within porous media.

The authors of \cite{Ped01} derived the governing equations for $\langle k^{\mathrm{usual}} \rangle$ and $\langle \varepsilon^{\mathrm{usual}} \rangle$. Their framework, which was used by other studies \cite{Brag, ChanH07}, is also employed in our study as follows:
\begin{eqnarray}
&\rho& \left[ \frac{\partial}{\partial t} (\phi \langle k^{\mathrm{usual}} \rangle) + \nabla \cdot \left( \bm{u}_D \langle k^{\mathrm{usual}} \rangle \right) \right] \nonumber \\
&=& \nabla \cdot \left[ \rho \left( \nu +\frac{\nu_T}{\sigma_k} \right) \nabla (\phi \langle k^{\mathrm{usual}} \rangle) \right]-\rho \phi \langle \overline{ \bm{u}' \bm{u}' } \rangle : \nabla \bm{u}_D \nonumber \\
& &+c_k \rho \phi \frac{\langle k^{\mathrm{usual}} \rangle |\bm{u}_D|}{\sqrt{K}} -\rho \phi \langle \varepsilon^{\mathrm{usual}} \rangle, \label{eqk1}
\end{eqnarray}
\begin{eqnarray}
&\rho& \left[ \frac{\partial}{\partial t} (\phi \langle \varepsilon^{\mathrm{usual}} \rangle) + \nabla \cdot \left( \bm{u}_D \langle \varepsilon^{\mathrm{usual}} \rangle \right) \right] \nonumber \\
&=& \nabla \cdot \left[ \rho \left( \nu +\frac{\nu_T}{\sigma_{\varepsilon}} \right) \nabla (\phi \langle \varepsilon^{\mathrm{usual}} \rangle) \right] -C_{\varepsilon 1} ( \rho \phi \langle \overline{ \bm{u}' \bm{u}' } \rangle : \nabla \bm{u}_D) \frac{\langle \varepsilon^{\mathrm{usual}} \rangle}{\langle k^{\mathrm{usual}} \rangle}  \nonumber \\
& &+C_{\varepsilon 2} \rho \phi \left(c_k \frac{\langle \varepsilon^{\mathrm{usual}} \rangle |\bm{u}_D|}{\sqrt{K}} -\frac{\langle \varepsilon^{\mathrm{usual}} \rangle^2}{\langle k^{\mathrm{usual}} \rangle} \right), \label{eqe1}
\end{eqnarray}
where $K$ denotes the permeability of the porous medium, $\bm{u}_D \equiv \phi \langle \overline{\bm{u}} \rangle$, $\langle \overline{\bm{u}' \bm{u}' } \rangle : \nabla \bm{u}_D \equiv \langle \overline{u'_i u'_j} \rangle \partial (u_D)_i /\partial x_j$, and $\sigma_k$, $\sigma_{\varepsilon}$, $c_k$, $C_{\varepsilon 1}$, and $C_{\varepsilon 2}$ are model parameters. The velocity field $\bm{u}_D$ and permeability are related as
\begin{equation}
\bm{u}_D=-\frac{\rho K}{\nu} \nabla \langle \overline{p} \rangle, \label{eqDar}
\end{equation}
which is the well-known Darcy law, with $p$ denoting the fluid pressure. Hereinafter, we will refer to $\bm{u}_D$ as the Darcy flow and the velocity field $\langle \bm{u}' \rangle$ as turbulence. Finally, we must define the eddy viscosity $\nu_T$ to close the macroscopic governing equation system \cite{Ped01} as
\begin{equation}
\nu_T=C_\mu \frac{\langle k^{\mathrm{usual}} \rangle^2}{\langle \varepsilon^{\mathrm{usual}} \rangle}, \label{eqEV}
\end{equation}
where $C_\mu$ is an empirical constant.

The first terms in the right-hand sides of Eqs. (\ref{eqk1}) and (\ref{eqe1}) describe the diffusion effects of $\langle k^{\mathrm{usual}} \rangle$ and $\langle \varepsilon^{\mathrm{usual}} \rangle$, respectively. The second terms, including the Reynolds stress, are production terms, which express the energy conveyed from the Darcy flow into the turbulence. The terms $-\rho \phi \langle \varepsilon^{\mathrm{usual}} \rangle$ and $-C_{\varepsilon 2} \rho \phi \langle \varepsilon^{\mathrm{usual}} \rangle^2/\langle k^{\mathrm{usual}} \rangle$ denote the energy dissipation due to the fluid viscosity. Such terms also appear in the governing equations for the usual turbulent flow with the limit $\phi \to 1$. Additionally, other production terms in porous media are introduced, such as the terms including $K$ in Eqs. (\ref{eqk1}) and (\ref{eqe1}). Notably, $1/\sqrt{K}$ is a measure of the flow resistance [see Eq. (\ref{eqDar})], and the higher kinetic energy of $\bm{u}_D$ will be converted to $\langle k^{\mathrm{usual}} \rangle$ with higher $1/\sqrt{K}$ via the pore walls. Noteworthily, the usual turbulent flow vanishes at the steady state for homogeneous $\bm{u}_D$ because the term with $1/\sqrt{K}$ does not exist. For porous media, the turbulence can continue to remain in such a case because the energy is always supplied from the Darcy flow into the turbulence via the pore walls, which plays an important role in this study.

For the usual turbulent flow, the parameter values are widely assumed as $C_{\mu}=0.09, \ \sigma_k=1.0, \ C_{\varepsilon 1}=1.44, \ C_{\varepsilon 2}=1.9-1.92$, and $\sigma_\varepsilon=1.2-1.3$ \cite{Lau74, Shi, Fen} and are considered universal, i.e., they do not depend on the flow. We will use these usual parameter values ($C_{\varepsilon 2}$ and $\sigma_{\varepsilon}$ are assumed to be 1.9 and 1.2, respectively) assuming that the flow is close to the usual turbulent flow as the first approximation \cite{Get, Nak08}. Additionally, we assume $c_k=0.28$, which was numerically derived by solving the Navier$-$Stokes equation for the flow through arrays of circular rods and by imposing the condition that the obtained results can be reproduced by using the macroscopic equations for $\langle k^{\mathrm{usual}} \rangle$ and $\langle \varepsilon^{\mathrm{usual}} \rangle$ \cite{Ped01, Chand06, Nim}.

However, obtaining the value of $c_k$ is somewhat difficult, and thus some researchers insist other values, e.g.,  $c_k=0.16-0.34$ \cite{Nak99}, and $c_k=0.22-0.34$ \cite{Chand06}. These differences are based on the porosity and configurations of pores. Nonetheless, the following discussion does not qualitatively change while choosing values in the range suggested for $c_k$ by previous studies. This is because the solid phase is assumed to be rigid and the porosity and the pore configurations remain unchanged, thereby indicating that parameter $c_k$ does not change temporally. 

Finally, we consider that $\langle k^{\mathrm{usual}} \rangle$ and $\langle \varepsilon^{\mathrm{usual}} \rangle$ correspond to $k$ and $\varepsilon$ in porous media, respectively. Therefore, we will express $\langle k^{\mathrm{usual}} \rangle$ and $\langle \varepsilon^{\mathrm{usual}} \rangle$ as $k$ and $\varepsilon$, respectively, in the following discussion. Equations (\ref{eqk1}) and (\ref{eqe1}) are considered to be the governing equations for $k$ and $\varepsilon$.



\section{Homogeneous System and Analytical Treatments} \label{secHSAT}

\subsection{Nullclines and Solution Orbit} \label{secNS}

\begin{figure}[tbp]
\centering
\includegraphics[width=8.5cm]{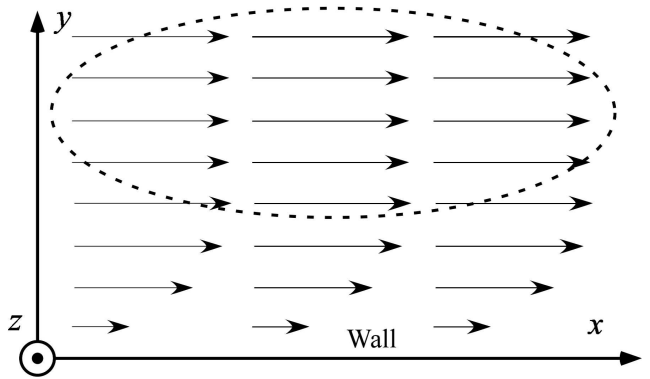}
\caption{Constant Darcy flow. We consider here a semi-infinite porous medium on an impermeable substrate. The substrate is assumed to be located on the $x-z$ plane and functions as a wall for the porous medium. The arrows represent the Darcy flow field, with the positive $x$ axis denoting the flow direction. The flow within the broken circle can be considered a homogeneous, constant Darcy flow because the region is far from the wall, and thus the effect of the wall on the flow can be considered negligible.}
\label{FigMS}
\end{figure}

We assume a homogeneous isotropic porous medium, and the constant Darcy flow, which is defined as a flow whose velocity is constant everywhere. This assumption has been considered for both the usual turbulent flow \cite{Yak, Bes, Zhao} and turbulent flow in porous media \cite{Mas, Nak99, Ped01}. For example, Zhao \cite{Zhao} insisted that such a situation is realized in the region outside the boundary layer over a flat plate, as shown in Fig. \ref{FigMS}. Here, we choose the $x$ axis along the Darcy flow direction. We can consider that a constant fluid pressure gradient is maintained along the $x$ axis by providing fluid from an infinitely far point. It is also important that $\bm{u}_D$ can be written as $\bm{u}_D=u_D \hat{\bm{x}}$, where $u_D$ denotes a positive constant and $\hat{\bm{x}}$ a unit vector directed along the positive $x$ axis. The turbulence is assumed to be homogeneous and isotropic. We also assume $\rho$ and $\phi$ to be constant, as mentioned in Sect. \ref{secMS}.

Under this condition, we can neglect the spatial differentiations in governing Eqs. (\ref{eqk1}) and (\ref{eqe1}); therefore, we can express the governing equations for $k$ and $\varepsilon$ as
\begin{equation}
\frac{\partial k}{\partial t} =\frac{c_k u_D}{\sqrt{K}} k -\varepsilon, \label{eqkh}
\end{equation}
\begin{equation}
\frac{\partial \varepsilon}{\partial t} =C_{\varepsilon 2} \frac{\varepsilon}{k} \left( \frac{c_k u_D}{\sqrt{K}} k -\varepsilon \right), \label{eqeh}
\end{equation}
respectively. We should emphasize that exactly the same straight line given by the equation,
\begin{equation}
\varepsilon=\frac{c_k u_D}{\sqrt{K}}k \label{eqNC}
\end{equation}
is a nullcline common to both the variables on the $k-\varepsilon$ phase space. Hence, the solution orbits are neither horizontal nor vertical where the solution orbits and this nullcline intersect \cite{Suz17}. Moreover, this line is a line attractor or repeller, as observed in other systems including \cite{Suz17}. Furthermore, the straight line $\varepsilon=0$ ($k$ axis) is a nullcline for $\varepsilon$.

We now obtain the analytical forms of the solution orbits. Equations (\ref{eqkh}) and (\ref{eqeh}) yield
\begin{equation}
\frac{\partial \varepsilon}{\partial k}=C_{\varepsilon 2} \frac{\varepsilon}{k}. \label{eqke}
\end{equation}
We assume constant $k_0 \neq 0$ and $\varepsilon_0 \neq 0$, where $k_0$ and $\varepsilon_0$ are the initial values of $k$ and $\varepsilon$, respectively. Accordingly, Eq. (\ref{eqke}) easily leads to the solution as follows: 
\begin{equation}
\varepsilon=\varepsilon_0 \left( \frac{k}{k_0} \right)^{C_{\varepsilon 2}}. \label{eqSO}
\end{equation}

\begin{figure}[tbp]
\centering
\includegraphics[width=8.5cm]{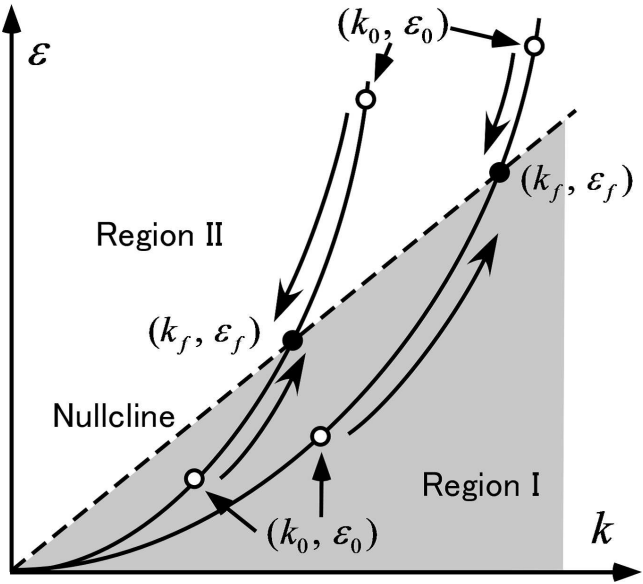}
\caption{Direction of the temporal evolutions of variables in the $k-\varepsilon$ phase space. The black curves represent the solution orbits. The broken straight line describes the nullcline. The shaded area is Region I, and the white area is Region II. Four cases for $(k_0, \varepsilon_0)$ are exemplified.}
\label{FigDTE}
\end{figure}

We now show that the common nullcline $\varepsilon=c_k u_D k/\sqrt{K}$ is a line attractor. Figure \ref{FigDTE} shows the direction of the temporal evolutions of $k$ and $\varepsilon$. First, we define Region I as the region $0<\varepsilon < c_k u_D k/\sqrt{K}$ and Region II as the region $\varepsilon > c_k u_D k/\sqrt{K}>0$, in the $k-\varepsilon$ phase space. We also define $(k_f, \varepsilon_f)$ as the point where the solution orbit (\ref{eqSO}) intersects the nullcline. With these definitions in hand, we obtain the relations $k_0<k_f$ and $\varepsilon_0<\varepsilon_f$ if $(k_0, \varepsilon_0)$ is in Region I because $\varepsilon \propto k$ on the nullcline, $\varepsilon \propto k^{C_{\varepsilon 2}}$ on the solution orbit, and $C_{\varepsilon 2} >1$. We can also conclude that $k_0>k_f$ and $\varepsilon_0>\varepsilon_f$ if $(k_0, \varepsilon_0)$ is in Region II. Second, we emphasize that $k$ and $\varepsilon$ increase (decrease) with increasing time in Region I (II), as the right-hand sides of Eqs. (\ref{eqkh}) and (\ref{eqeh}) are positive (negative). Therefore, if $(k_0, \varepsilon_0)$ is in Region I (II), the solution moves to the upper right (lower left) and gets absorbed into the point $(k_f, \varepsilon_f)$ on the nullcline with the limit $t \to \infty$ (see Fig. \ref{FigDTE}). Thus, we can conclude that the nullcline is a line attractor, not a repeller. The steady stable solution is given by $(k, \varepsilon)=(k_f, \varepsilon_f)$, and this state will be referred to as the final state in the following discussions. Noteworthily, $k$ and $\varepsilon$ vanish with the limit $t \to \infty$ for the usual turbulent flow, even though solution orbit (\ref{eqSO}) is valid for such flows \cite{Zhao}. Actually, the usual turbulent flow is described by the limit $K \to \infty$. The nullcline is the $k$ axis in such a case, with Region I vanishing. Therefore, all the solutions are absorbed into the origin. The terms including the finite $K$ enable the homogeneous isotropic turbulence to survive as $t \to \infty$ under the constant Darcy flow.

Therefore, the relationship $\varepsilon_f=c_k u_D k_f/\sqrt{K}$ must be satisfied for the constant Darcy flow and fully developed homogeneous isotropic turbulence. Using this relationship and Eq. (\ref{eqSO}), $k_f$ and $\varepsilon_f$ are given as
\begin{equation}
k_f =\left( \frac{c_k u_D}{\sqrt{K}} \right)^{1/(C_{\varepsilon 2}-1)} \left( \frac{k_0^{C_{\varepsilon 2}}}{\varepsilon_0} \right)^{1/(C_{\varepsilon 2}-1)}, \label{eqkf}
\end{equation}
and
\begin{equation}
\varepsilon_f =\left( \frac{c_k u_D}{\sqrt{K}} \right)^{C_{\varepsilon 2}/(C_{\varepsilon 2}-1)} \left( \frac{k_0^{C_{\varepsilon 2}}}{\varepsilon_0} \right)^{1/(C_{\varepsilon 2}-1)}, \label{eqef}
\end{equation}
respectively. These values are important in investigating the temporal evolution of the eddy viscosity, as discussed in Sect. \ref{secTEEV}.

\subsection{Temporal Evolution of Eddy Viscosity} \label{secTEEV}

We now investigate the temporal evolution of the eddy viscosity. We note that the initial value of $\nu_T$, $\nu_{T0}$, is clearly given by
\begin{equation}
\nu_{T0} =C_\mu \frac{k_0^2}{\varepsilon_0}, \label{eqnt0}
\end{equation}
and the value of $\nu_T$ at the final state, $\nu_{Tf}$, can be expressed as
\begin{equation}
\nu_{Tf}=C_\mu \left( \frac{c_k u_D}{\sqrt{K}} \right)^{-(C_{\varepsilon 2}-2)/(C_{\varepsilon 2}-1)} \left( \frac{k_0^{C_{\varepsilon 2}}}{\varepsilon_0} \right)^{1/(C_{\varepsilon 2}-1)}. \label{eqntf}
\end{equation}
This value is derived from Eqs. (\ref{eqkf}) and (\ref{eqef}). The ratio between the final and initial viscosities is given by
\begin{equation}
\frac{\nu_{Tf}}{\nu_{T0}} =\left( \frac{c_k u_D}{\alpha \sqrt{K}} \right)^{-(C_{\varepsilon 2} -2)/(C_{\varepsilon 2}-1)},
\end{equation}
where $\alpha \equiv \varepsilon_0/k_0$. Here, we emphasize $-(C_{\varepsilon 2}-2)/(C_{\varepsilon 2}-1) >0$, as $C_{\varepsilon 2}=1.9$. Let us assume that $(k_0, \varepsilon_0)$ is in Region I. In this case, $c_k u_D/\sqrt{K} > \alpha$, and hence $\nu_{Tf} > \nu_{T0}$. Alternatively, if we assume that $(k_0, \varepsilon_0)$ is in Region II, then $\nu_{Tf} < \nu_{T0}$. Briefly, if $(k_0, \varepsilon_0)$ is in Region I, the final disturbance is stronger than the initial one, and if $(k_0, \varepsilon_0)$ is in Region II, the final disturbance is weaker than the initial one. Notably, $\nu_T$ is a quantitative measure of the disturbance intensity, as mentioned in Sect. \ref{secNS}.

From a physical viewpoint, the energy supply from the Darcy flow to the turbulence is dominant in Region I, and energy dissipation is dominant in Region II. To elucidate on this, let us consider the straight line $k=\mathrm{const.}$ On the line, $\varepsilon$ is greater in Region II than in Region I. Moreover, as $\varepsilon$ denotes the energy dissipation, its effect dominates and consequently the disturbance is reduced in Region II.

\begin{figure}[tbp]
\centering
\includegraphics[width=8.5cm]{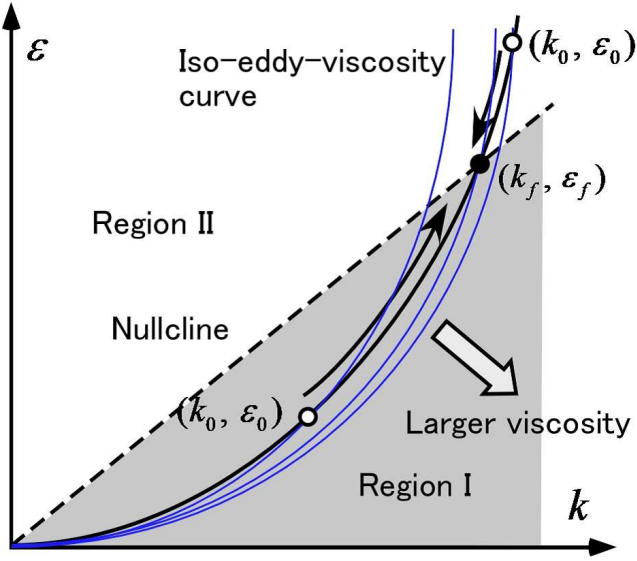}
\caption{Temporal evolution of the eddy viscosity. The thin blue solid lines represent the iso-eddy-viscosity curves. The black curves, broken straight line, and the shaded and white regions are the same as those in Fig. \ref{FigDTE}. (Color online)}
\label{FigEV}
\end{figure}

The investigation above can be easily understood on the basis of $k-\varepsilon$ phase space. We now investigate the temporal evolution of the eddy viscosity. We begin by obtaining the analytical form of the curve on which the eddy viscosity takes the same value (iso-eddy-viscosity curve) on the $k-\varepsilon$ phase space. From Eq. (\ref{eqEV}), we have
\begin{equation}
\varepsilon=\frac{C_\mu}{\nu_T} k^2 \label{eqIV}
\end{equation}
for the curve. This equation insists that the iso-eddy-viscosity curve is a parabola whose vertex lies at the origin in the $k-\varepsilon$ phase space. Comparing Eqs. (\ref{eqSO}) and (\ref{eqIV}), we can show that if $(k_0, \varepsilon_0)$ is in Region I, then the eddy viscosity increases with time. This increase in the eddy viscosity occurs because the gradient of the solution orbit is smaller than that of curve (\ref{eqIV}) at their crossing point. Notably, $C_{\varepsilon 2}=1.9<2$ (see Fig. \ref{FigEV}). With increase in time, the solution migrates to the iso-eddy-viscosity curve with a higher eddy viscosity than before. For the same reason, if $(k_0, \varepsilon_0)$ is in Region II, $\nu_T$ decreases with time (see Fig. \ref{FigEV}).

\subsection{Relationship between $k_0$ and $\varepsilon_0$ and Universal Power Law} \label{secrke}

We allowed any values for $k_0$ and $\varepsilon_0$ in the above discussion because, particularly, determining $\varepsilon_0$ is an arduous task. However, several previous studies assumed that the values of $k_0$ and $\varepsilon_0$ are ruled by some simple relations. For example, some researchers assumed that the ratio $\nu_T/\nu$ is constant for the usual turbulent flow \cite{Lin, Bos} and for the turbulent flow in porous media \cite{Hsu}, leading to the relation $\varepsilon_0 \propto k_0^2$. Another example $\varepsilon_0 \propto k_0$ was used for the usual turbulent flow \cite{Yak} and for the turbulent flow in porous media \cite{Jac}. The relation $\varepsilon_0 \propto k_0^{3/2}$ was adopted for the usual turbulent flow \cite{Yos}. Although several ideas were considered, we, at least, require ``the larger $\varepsilon_0$ with larger $k_0$.'' Accordingly, we assume $\varepsilon_0 \propto k_0^n$, where $n$ denotes a positive number that satisfies the condition $1 \le n \le 2$.

We first analytically show the dependence of $\nu_{T0}$ and $\nu_{Tf}$ on $k_0$. We define $\beta$ as the coefficient of proportionality for $\varepsilon_0$ and $k_0^n$, i.e., $\varepsilon_0 =\beta k_0^n$. From this definition and Eqs. (\ref{eqnt0}) and (\ref{eqntf}), we obtain the relations
\begin{equation}
\nu_{T0} =C_\mu \frac{k_0^{2-n}}{\beta}, \label{eqnt0k0-n} 
\end{equation}
and
\begin{eqnarray}
\nu_{Tf}&=&C_\mu \left( \frac{c_k u_D}{\sqrt{K}} \right)^{-(C_{\varepsilon 2}-2)/(C_{\varepsilon 2}-1)} \left( \frac{k_0^{C_{\varepsilon 2}-n}}{\beta} \right)^{1/(C_{\varepsilon 2}-1)}. \label{eqntfk0-n} 
\end{eqnarray}

Furthermore, if $n \neq 1$, we have the fixed point $(k_{0,n}^{\mathrm{fix}}, \varepsilon_{0,n}^{\mathrm{fix}})$ on the $k-\varepsilon$ phase space. This point does not move with increase in time. The point $(k_{0,n}^{\mathrm{fix}}, \varepsilon_{0,n}^{\mathrm{fix}})$ is the point of intersection of the nullcline and the curve that describes the relationship between $k_0$ and $\varepsilon_0$. From this definition, $k_{0,n}^{\mathrm{fix}}$ is given by the solution of the equation
\begin{equation}
\frac{c_k u_D}{\sqrt{K}} k =\beta k^n,
\end{equation}
which reads
\begin{equation}
k_{0,n}^{\mathrm{fix}}=\left( \frac{c_k u_D}{\beta \sqrt{K}} \right)^{1/(n-1)}.
\end{equation}
We can also obtain
\begin{equation}
\varepsilon_{0,n}^{\mathrm{fix}}=\beta (k_{0,n}^{\mathrm{fix}})^n=\beta^{-1/(n-1)} \left( \frac{c_k u_D}{\sqrt{K}} \right)^{n/(n-1)}.
\end{equation}
With these values, the fixed eddy viscosity, $\nu_{T,n}^{\mathrm{fix}}$, is given as
\begin{equation}
\nu_{T,n}^{\mathrm{fix}} \equiv C_\mu \frac{(k_{0,n}^{\mathrm{fix}})^2}{\varepsilon_{0,n}^{\mathrm{fix}}}=C_\mu \beta^{-1/(n-1)} \left( \frac{c_k u_D}{\sqrt{K}} \right)^{(2-n)/(n-1)}. \label{eqntfix-n}
\end{equation}

We now investigate the relationship between the normalized values $\tilde{\nu}_{T0} \equiv \nu_{T0}/\nu_{T,n}^{\mathrm{fix}}$ and $\tilde{\nu}_{Tf} \equiv \nu_{Tf}/\nu_{T,n}^{\mathrm{fix}}$ on the basis of Eqs. (\ref{eqnt0k0-n}), (\ref{eqntfk0-n}), and (\ref{eqntfix-n}) with $n \neq 1$ and $2$. From these equations, we can obtain the relation
\begin{equation}
\tilde{\nu}_{Tf} = (\tilde{\nu}_{T0})^{(C_{\varepsilon 2}-n)/[(2-n)(C_{\varepsilon 2}-1)]}. \label{eqntf0-a}
\end{equation}
Using $C_{\varepsilon 2}=1.9$ and $1<n<2$, we can prove that $(C_{\varepsilon 2}-n)/[(2-n)(C_{\varepsilon 2}-1)]$ is positive when $1<n<C_{\varepsilon 2}$. Therefore, for $1<n<C_{\varepsilon 2}$, $\tilde{\nu}_{Tf}$ monotonically increases with increasing $\tilde{\nu}_{T0}$; i.e., the stronger initial disturbance leads to the stronger final disturbance when $1<n<C_{\varepsilon 2}$. On the other hand, for $C_{\varepsilon 2}<n<2$, $\tilde{\nu}_{Tf}$ monotonically decreases with increasing $\tilde{\nu}_{T0}$; i.e., the stronger initial disturbance leads to the weaker final disturbance when $C_{\varepsilon 2}<n<2$. If $n=C_{\varepsilon 2}$, $\tilde{\nu}_{Tf}$ is unity ($\nu_{Tf}=\nu_{T,n}^{\mathrm{fix}}$), which is independent of $\tilde{\nu}_{T0}$. With the $k-\varepsilon$ phase space, we next show the sensitive dependence of $\tilde{\nu}_{Tf}$ on the initial conditions for the cases $C_{\varepsilon 2}<n<2$ and $1<n<C_{\varepsilon 2}$. For the cases $n=1$ and $n=2$, we investigate such relations by employing Eqs. (\ref{eqnt0k0-n}) and (\ref{eqntfk0-n}).

\begin{figure}[tbp]
\centering
\includegraphics[width=8.5cm]{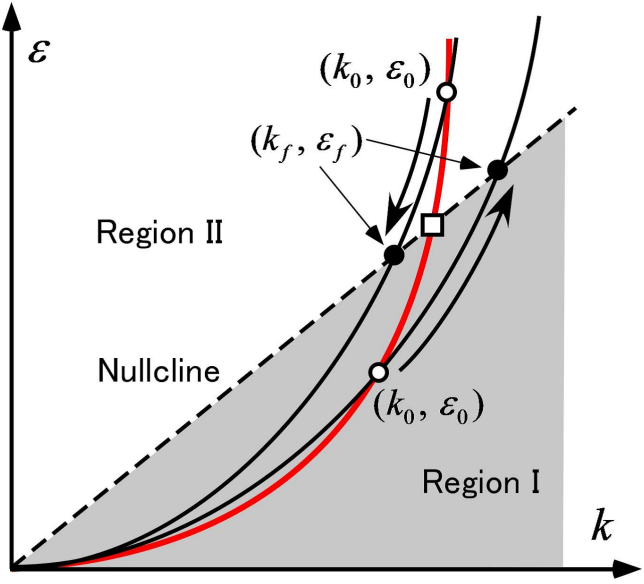}
\caption{Initial and final values of $k$ and $\varepsilon$ for the case $C_{\varepsilon 2} <n<2$. The red solid line represents the curve on which the point $(k_0, \varepsilon_0)$ exists. The white square describes the fixed point $(k_{0,n}^{\mathrm{fix}}, \varepsilon_{0,n}^{\mathrm{fix}})$. The black curves, broken straight line, and the shaded and white regions are the same as those in Fig. \ref{FigDTE}. (Color online)}
\label{FigInit1}
\end{figure}

We visualize the analytical result by using the $k-\varepsilon$ phase space for the case $C_{\varepsilon 2}<n<2$. The solution orbits and the curve that describes the relationship $\varepsilon_0=\beta k_0^n$ are shown in Fig. \ref{FigInit1}. Notably, on the nullcline and red curve, $\nu_T$ is higher on the points farther from the origin [see Eq. (\ref{eqIV})]. Additionally, a point nearer to the origin on the red curve evolves to a point farther from the origin on the nullcline with the limit $t \to \infty$ because the power of the solution orbit ($C_{\varepsilon 2}$) is smaller than $n$. We can, thus, conclude that a lower $\tilde{\nu}_{T0}$ results in a higher $\tilde{\nu}_{Tf}$. The value $\tilde{\nu}_{T0}=1$ is critical and fixed; i.e., if $\tilde{\nu}_{T0}=1$, then $\tilde{\nu}_{Tf}=1$.

Moreover, the power law (\ref{eqntf0-a}) implies the sensitivity of the system behavior within $C_{\varepsilon 2} <n<2$. Let us assume that $(k_0, \varepsilon_0)$ is in Region I and $\tilde{\nu}_{T0} \ll 1$. Here, note that $\tilde{\nu}_{Tf}$ is described by the negative power of $\tilde{\nu}_{T0}$, and that a significantly high $\tilde{\nu}_{Tf}$ is induced by such a negligible value of $\tilde{\nu}_{T0}$. This sensitivity of the final state to the initial condition is peculiar to a turbulent flow in porous media. Although the usual turbulent flow is also sensitive to the initial condition, the sensitivity characterized by the power law (\ref{eqntf0-a}) is specific to the porous media. This law is universal because the power $(C_{\varepsilon 2}-n)/[(2-n)(C_{\varepsilon 2}-1)]$ does not depend on the media details including porosity and permeability. 

\begin{figure}[tbp]
\centering
\includegraphics[width=8.5cm]{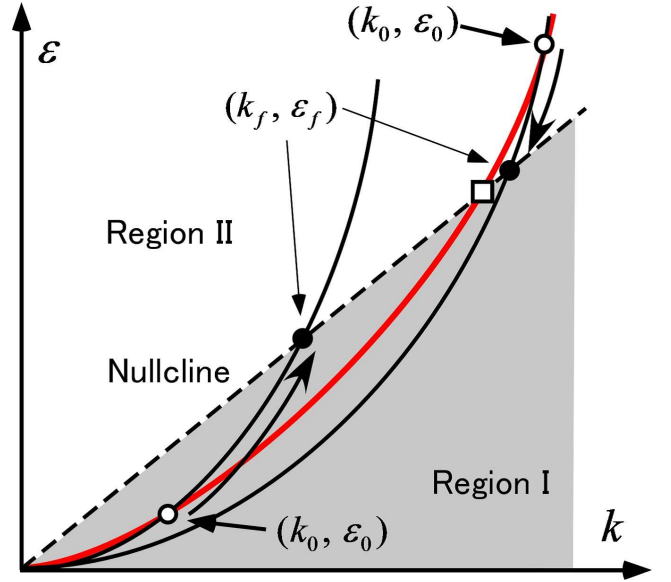}
\caption{Initial and final values of $k$ and $\varepsilon$ for the case $1<n<C_{\varepsilon 2}$. The red solid line represents the curve on which the point $(k_0, \varepsilon_0)$ exists. The white square describes the fixed point $(k_{0,n}^{\mathrm{fix}}, \varepsilon_{0,n}^{\mathrm{fix}})$. The black curves, broken straight line, and the shaded and white regions are the same as those in Fig. \ref{FigDTE}. (Color online)}
\label{FigInit2}
\end{figure}

We then consider the case $1<n<C_{\varepsilon 2}$. We have visualized the behavior of the solution for this case in Fig. \ref{FigInit2}, where the solution orbits and the curve that describes the relationship $\varepsilon_0=\beta k_0^n$ are illustrated. On the nullcline and red curve, as previously mentioned, $\nu_T$ is higher on the point farther from the origin. Additionally, a point nearer to the origin on the red curve evolves to a point nearer to the origin on the nullcline with the limit $t \to \infty$ because the gradient of the solution orbit is larger than that of the red curve where both curves intersect (Notably, $n<C_{\varepsilon 2}$). In such a case, a higher $\tilde{\nu}_{T0}$ results in a higher $\tilde{\nu}_{Tf}$. Nonetheless, $\tilde{\nu}_{T0}=\tilde{\nu}_{Tf}=1$ remains a fixed value, as observed for the case $C_{\varepsilon 2}<n<2$.

Additionally, we can show that the universal sensitivity reappears. Because the power $(C_{\varepsilon 2}-n)/[(2-n)(C_{\varepsilon 2}-1)]$ is positive and less than unity for $1<n<C_{\varepsilon 2}$, $\tilde{\nu}_{Tf}$ can be significantly higher than $\tilde{\nu}_{T0}$ with $\tilde{\nu}_{T0} \ll 1$. Nonetheless, note that the sensitivities for the cases $C_{\varepsilon 2}<n<2$ and $1<n<C_{\varepsilon 2}$ are qualitatively different from each other. To clarify this, let us consider the limit $\tilde{\nu}_{T0} \to 0$. With this limit, we have $\tilde{\nu}_{Tf} \to \infty$ for the case $C_{\varepsilon 2}<n<2$ and $\tilde{\nu}_{Tf} \to 0$ for the case $1<n<C_{\varepsilon 2}$. We can insist that the sensitivity discontinuously changes for the cases $C_{\varepsilon 2}<n<2$ and $1<n<C_{\varepsilon 2}$, and that the sensitivity is stronger in the former case than in the latter one. Finally, the common nullcline (\ref{eqNC}) is important for the universal sensitivity. The sensitivity due to the common nullcline has been shown to emerge in a dynamic earthquake slip process, as detailed in \cite{Suz17}.

If $n=1$, we cannot define the fixed point and cannot employ the power law (\ref{eqntf0-a}). In this case, we have a relationship $\nu_{Tf} \propto \nu_{T0}$ from Eqs. (\ref{eqnt0k0-n}) and (\ref{eqntfk0-n}). The sensitivity does not emerge when $n=1$.

If $n=2$, we can confirm that $\nu_{T0}$ has constant value $C_{\mu}/\beta$, and thus we cannot use the power law (\ref{eqntf0-a}). Alternatively, we can employ Eq. (\ref{eqntfk0-n}), which represents the power law, as
\begin{equation}
\nu_{Tf} \propto k_0^{(C_{\varepsilon 2} -2)/(C_{\varepsilon_2} -1)} \propto \varepsilon_0^{(C_{\varepsilon 2} -2)/[2(C_{\varepsilon_2} -1)]}.
\end{equation}
We can conclude that $\nu_{Tf}$ is sensitive to $k_0$ or $\varepsilon_0$ because $(C_{\varepsilon 2}-2)/(C_{\varepsilon 2}-1)$ is negative. 

Several previous calculations \cite{Nak99, Ped01} assumed a homogeneous Darcy flow and constant $k_0$ and $\varepsilon_0$, and maintained their values at the boundary of a one-dimensional computational domain with increasing time. In these calculations, the relationship $\nu_{T0}>\nu_{Tf}$ was demonstrated wherein the distance from the boundary was so large that the assumption of the homogeneous isotropic turbulence held valid. However, the relationship $\nu_{T0}<\nu_{Tf}$ holds true in laboratory experiments such as the one in \cite{Nak07}. A finite $\nu_{Tf}$ is generated from the negligible $\nu_{T0}$ value in the experiments. This variation observed in previous studies may be explained using the single framework constructed in this study.





\subsection{General Treatment of Difference in the Initial Conditions and Its Application to Nonlinear Mathematics} \label{secGA}

If the condition $\varepsilon_0 \propto k_0^n$ is not assumed, we cannot derive relations (\ref{eqnt0k0-n}) through (\ref{eqntf0-a}). However, let us consider two points $(k_0, \varepsilon_0)$ and $(k_0 +\delta k_0, \varepsilon_0 +\delta \varepsilon_0)$, where $\delta k_0$ and $\delta \varepsilon_0$ denote positive or negative amounts that satisfy the conditions $|\delta k_0/k_0| \ll 1$ and $|\delta \varepsilon_0/\varepsilon_0| \ll 1$, respectively, on the $k-\varepsilon$ phase space. Actually, $\nu_{Tf}$ can be sensitive to $\delta k_0$ and $\delta \varepsilon_0$, and the sensitivity should be analytically clarified. Note that the sensitivity investigated above was that of $\nu_{Tf}$ on $\nu_{T0}$, while the sensitivity studied here is that of the difference in $\nu_{Tf}$ on the difference in $\nu_{T0}$.

To clarify the sensitivity, we consider the following two relationships:
\begin{eqnarray}
&\nu_{T0}&(k_0 +\delta k_0, \varepsilon_0 +\delta \varepsilon_0)-\nu_{T0}(k_0, \varepsilon_0) \nonumber \\
=&\nu_{T0}&(k_0, \varepsilon_0) \left( \frac{2}{k_0} \delta k_0 -\frac{1}{\varepsilon_0} \delta \varepsilon_0 +O(\delta k_0^2)+O(\delta \varepsilon_0^2) \right),
\end{eqnarray}
and
\begin{eqnarray}
&\nu_{Tf}&(k_0 +\delta k_0, \varepsilon_0 +\delta \varepsilon_0)-\nu_{Tf}(k_0, \varepsilon_0) \nonumber \\
&=&\nu_{Tf}(k_0, \varepsilon_0) \times \nonumber \\
& &\left( \frac{C_{\varepsilon 2}}{C_{\varepsilon 2}-1}\frac{1}{k_0} \delta k_0 -\frac{1}{C_{\varepsilon 2}-1} \frac{1}{\varepsilon_0} \delta \varepsilon_0 +O(\delta k_0^2)+O(\delta \varepsilon_0^2) \right).
\end{eqnarray}
Therefore, if we neglect the terms of orders $\delta k_0^2$ and $\delta \varepsilon_0^2$ and higher, we obtain
\begin{eqnarray}
\frac{\nu_{Tf}(k_0 +\delta k_0, \varepsilon_0 +\delta \varepsilon_0)-\nu_{Tf}(k_0, \varepsilon_0)}{\nu_{T0}(k_0 +\delta k_0, \varepsilon_0 +\delta \varepsilon_0)-\nu_{T0}(k_0, \varepsilon_0)} \nonumber \\
=\frac{1}{C_{\varepsilon 2}-1} \frac{\nu_{Tf}(k_0, \varepsilon_0)}{\nu_{T0}(k_0, \varepsilon_0)}\left( 1+\frac{\varepsilon_0 (2-C_{\varepsilon 2})}{k_0 g -2 \varepsilon_0} \right),
\end{eqnarray}
where $g \equiv \delta \varepsilon_0/\delta k_0$.

\begin{figure}[tbp]
\centering
\includegraphics[width=8.5cm]{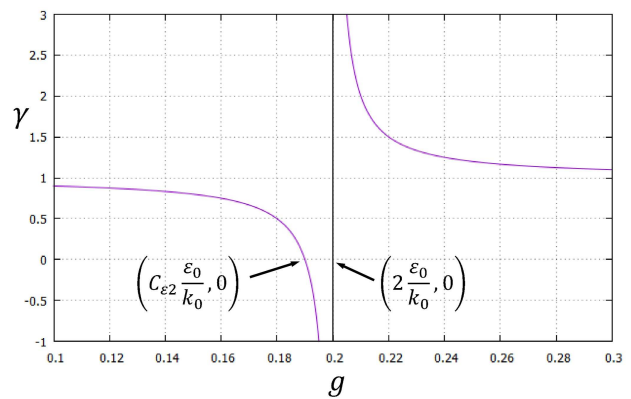}
\caption{Dependence of $\gamma$ on $g$. The black straight line represents $g=2 \varepsilon_0/k_0$. The values $k_0=1$ and $\varepsilon_0=0.1$ are employed. (Color online)}
\label{FigGamma}
\end{figure}

We refer to $1+\varepsilon_0 (2-C_{\varepsilon 2})/(k_0 g-2 \varepsilon_0)$ as $\gamma$, which is plotted in Fig. \ref{FigGamma} as a function of $g$. If $g$ is near $2 \varepsilon_0/k_0$, the slight disturbance in the initial condition is enlarged in $\nu_{Tf}$, thereby indicating strong sensitivity. Note that $\gamma$ is not a power, but the relative magnitude of the difference in $\nu_{Tf}$ to the difference in $\nu_{T0}$. Therefore, the sign of $\gamma$ does not determine the strength of the sensitivity, and thus only the absolute value of $\gamma$ is relevant.

The important value $g=2 \varepsilon_0/k_0$ has a physical meaning. Considering the condition $g=\delta \varepsilon_0/\delta k_0=2 \varepsilon_0/k_0$ and the limit $\delta k_0 \to 0$ and $\delta \varepsilon_0 \to 0$, the two points $(k_0, \varepsilon_0)$ and $(k_0+\delta k_0, \varepsilon_0+\delta \varepsilon_0)$ lie on the same iso-eddy-viscosity curve because $\varepsilon_0 \propto k_0^2$ on the curve; hence, the values of $\nu_{T0}(k_0, \varepsilon_0)$ and $\nu_{T0}(k_0+\delta k_0, \varepsilon_0+\delta \varepsilon_0)$ are exactly the same. Therefore, the difference in $\nu_{T0}$ can arbitrarily increase in the difference in $\nu_{Tf}$ when $g$ is close to $2 \varepsilon_0/k_0$.

\section{Discussion AND Conclusions} \label{secDisCon}

For the turbulent flow through porous media, we applied the $k-\varepsilon$ model and assumed the constant Darcy flow and the homogeneous isotropic turbulence. The eddy viscosities of the initial and final states are related via a power law, which is universal and independent of the media details including porosity and permeability. Particularly, the sign of the power depends on the initial condition; if $\varepsilon_0 \propto k_0^n$ is satisfied, it is negative (positive) for the case $C_{\varepsilon 2}<n<2$ ($1<n<C_{\varepsilon 2}$). For both the cases, the final eddy viscosity is sensitive to its initial value and the sensitivity is stronger for the $C_{\varepsilon 2}<n<1$ than for $1<n<C_{\varepsilon 2}$ cases. Furthermore, even if $n=2$, the final eddy viscosity sensitively depends on the initial values $k_0$ and $\varepsilon_0$. The sensitivity mentioned here is not observed for the usual turbulent flow, but it originates from the common nullcline for $k$ and $\varepsilon$, as consistent with a previous study \cite{Suz17}. The importance of the common nullcline should be emphasized in nonlinear mathematics. Moreover, the difference observed in the final eddy viscosity is also sensitive to that observed in the initial eddy viscosity.

Let us now consider a spatially inhomogeneous flow. In this case, the spatial differentiations in Eqs. (\ref{eqk1}) and (\ref{eqe1}) do not vanish. In particular, the coefficients of the diffusion terms include $\nu_T$, which enhances the diffusions of $k$ and $\varepsilon$. Although this diffusion term is not considered in this study, we qualitatively employ the effect of disturbance enhancing the time-averaged velocity in this section as the first step toward extending the present results by considering seismological slip as an example.  Quantitatively considering this effect will be an important future work. 


As mentioned in Sect. \ref{secIn}, fault rocks can be considered porous media. Two semi-infinite porous media in contact will slide when they experience shear stress, thereby mimicking the fault motion. Moreover, the contact surface can be regarded as a fault zone. Particularly, the contact surface cannot be a mirror-like plane and is rough. Therefore, here we use the term ``fault zone'' instead of the term ``fault plane.'' Previous studies were able to model two effects, namely, thermal pressurization and dilatancy effects (both of which change the fluid pressure on the fault zone, $p_F$), into a single framework \cite{Ric06, Suz09, Sega, Suz14, Suz17}. Herein, we assume that $p_F$ is initially the same as the ambient fluid pressure. The thermal pressurization effect increases $p_F$, whereas the dilatancy effect reduces $p_F$. Moreover, $p_F$ and the slip velocity on the fault zone assume a linear relationship: an increase (decrease) in $p_F$ corresponds to an increase (decrease) in the slip velocity (for a detailed discussion on this, kindly see \cite{Sega, Suz14, Suz17}). Moreover, the fluid flow also affects $p_F$ and induces slip velocity change, as mentioned in Sect. \ref{secIn}.


Let us provide a rough estimation of $\mathrm{Re}_p$ during a dynamic earthquake slip process. Importantly, $\nu=3 \times 10^{-7} \ [\mathrm{m}^2/\mathrm{s}]$ for heated (near 373.15 [K]) water, and we can roughly estimate $d$ to be in the range of $0.1 - 1 \  [\mathrm{mm}]$, as suggested by gouge sizes for natural fault rocks \cite{Ben, Wib}. We should also evaluate the characteristic flow velocity $v$, which is arduous to calculate. Specifically, to evaluate $v$, we must introduce the fault-zone thickness $T_F$ and characteristic time scale (rise time) $D$ and consider the value $T_F/D$ as a measure of $v$. For subduction thrust faults, the order of mm$-$cm seems reasonable for $T_F$ \cite{Uji07, Uji08, Kam, Row}. Noteworthily, $D$ is on the order of $0.1 - 10\  [\mathrm{s}]$ for earthquakes with a moment magnitude scale in the range of $6-7$ \cite{Ols, JiC, Avo}. Therefore, we can roughly estimate $v$ as $T_F/D \sim 0.01 \ - 10 \ [\mathrm{cm/s}]$. The corresponding pore Reynolds number at these ranges is approximately $0.033-3.3 \times 10^2$, with the latter satisfying condition (\ref{eqRep}). Accordingly, we can conclude that the turbulent flow plays a more important role for earthquakes with larger gouges, thicker fault zones, and shorter rise times.

The increase in $\nu_T$ enhances the fluid inflow or outflow because effective viscosities, $\nu+\nu_T/\sigma_k$ and $\nu+\nu_T/\sigma_k$, include $\nu_T$, as shown in Eqs. (\ref{eqk1}) and (\ref{eqe1}). For ordinary earthquakes, the fluid inflow is considered to occur \cite{Suz14}, and if the inflow is enhanced, the slip velocity would increase further because the subsequent thermal pressurization would raise $p_F$, indicating decrease in the friction stress. Therefore, the acceleration is considered positive. The acceleration of the initial phase of the slip may be explained by the feedback proposed herein.

We finally extend the result obtained in Sect. \ref{secGA} from the viewpoint of the nonlinear mathematics. We consider two quantities, $x_1$ and $x_2$. We define a single scalar value $V(x_1, x_2)$ on the $x_1-x_2$ phase space, and $V$ is assumed to be proportional to $x_1^a x_2^b$, where $a$ and $b$ denote constant numbers. The initial values of $x_1$ and $x_2$ are denoted by $x_{10}$ and $x_{20}$, respectively. The common nullcline describing the continuous attractor is assumed to exist on the $x_1-x_2$ phase space. Actually, a continuous attractor is a straight line that crosses the origin in many cases \cite{Suz17, Zhan}, and this assumption is also employed here. Therefore, the relationship $x_{2f} \propto x_{1f}$ is satisfied. Additionally, in many cases, the solution orbits can be described by $x_2 \propto x_1^N$, where $N$ denotes a constant, and we also adopt this relation. With these assumptions, we now expect $V_0 \equiv V(x_{10}, x_{20}) \propto x_{10}^{M_0} x_{20}^{N_0}$ and $V_f \equiv V(x_{1f}, x_{2f}) \propto x_{10}^{M_f} x_{20}^{N_f}$, where $M_0$, $N_0$, $N_f$, and $M_f$ denote constant numbers. We consider the disturbances $\delta x_{10}$ and $\delta x_{20}$ in $x_{10}$ and $x_{20}$, respectively. We impose the conditions $|\delta x_{10}/x_{10}| \ll 1$ and $|\delta x_{20}/x_{20}| \ll 1$, and allow both positive and negative $\delta x_{10}$ and $\delta x_{20}$. With this situation, we have the relations
\begin{eqnarray}
&V_0&(x_{10} +\delta x_{10}, x_{20} +\delta x_{20})-V_0(x_{10}, x_{20}) \nonumber \\
=&V_0&(x_{10}, x_{20}) \times \nonumber \\
& &\left( \frac{M_0}{x_{10}} \delta x_{10} +\frac{N_0}{x_{20}} \delta x_{20} +O(\delta x_{10}^2)+O(\delta x_{20}^2) \right),
\end{eqnarray}
and
\begin{eqnarray}
&V_f&(x_{10} +\delta x_{10}, x_{20} +\delta x_{20})-V_f(x_{10}, x_{20}) \nonumber \\
=&V_f&(x_{10}, x_{20}) \times \nonumber \\
& &\left( \frac{M_f}{x_{10}} \delta x_{10} +\frac{N_f}{x_{20}} \delta x_{20} +O(\delta x_{10}^2)+O(\delta x_{20}^2) \right),
\end{eqnarray}
which results in
\begin{eqnarray}
\frac{V_f(x_{10} +\delta x_{10}, x_{20} +\delta x_{20})-V_f(x_{10}, x_{20})}{V_0(x_{10} +\delta x_{10}, x_{20} +\delta x_{20})-V_0(x_{10}, x_{20})} \nonumber \\
=\frac{V_f(x_{10}, x_{20})}{V_0 (x_{10}, x_{20})} \frac{N_f x_{10} g +M_f x_{20}}{N_0 x_{10} g +M_0 x_{20}},
\end{eqnarray}
where $g \equiv \delta x_{20}/\delta x_{10}$. If we define $\gamma$ as $(N_f x_{10} g +M_f x_{20})/(N_0 x_{10} g +M_0 x_{20})$ in this case, the value of $\gamma$ determines the sensitivity of $V$ on the initial condition.

Continuous attractors have drawn the attention of researchers in terms of, particularly, neural networks \cite{Sam, Seu, YuJ, Mil, Zhan}. For example, $x_i$ described the activity of neuron $i$ in the model of \cite{Zhan}. When we consider two neurons, the straight line $x_2=x_1$ is the common nullcline, and this case is known to generate $N=-1$ \cite{Zhan}. In their model, we can consider $x_2$ as an example of $V$. We then have $V_0=x_{20}$ and $V_f=\sqrt{x_{10} x_{20}}$. Therefore, we have $M_0=0, N_0=1, M_f=1/2$, and $N_f=1/2$ in this case, leading to $\gamma=(g x_{10}+x_{20})/2x_{10} g$. Therefore, if $g$ is negligibly small, a negligible but finite difference in the initial condition is enlarged in the final value. However, if $g$ is near $-x_{20}/x_{10}$, the enlargement is not observed. The parameter $\gamma$ can be a measure for the sensitivity of physical quantities on the initial conditions in a system with common nullclines.

\acknowledgments

The author would like to thank Professor de Lemos and Professor K. Saitoh for fruitful discussions with them. He was supported by JSPS KAKENHI Grant Numbers JP16H06478 and JP20K03771. This study was supported by the Ministry of Education, Culture, Sports, Science and Technology (MEXT) of Japan, under its The Second Earthquake and Volcano Hazards Observation and Research Program (Earthquake and Volcano Hazard Reduction Research). This study was also supported by ERI JURP 2018-G-04, 2019-G-02 and 2020-G-08. This work was partially supported by Aoyama Gakuin University research grant ``Ongoing Research Program.'' 

\color{red}

\renewcommand{\theequation}{A\arabic{equation}}
\setcounter{equation}{0}

\appendix

\color{black}



%

\end{document}